\documentclass[prb,twocolumn,aps,superscriptaddress,showpacs,preprintnumbers,amssymb]{revtex4-1}
\usepackage{graphicx}
\usepackage{dcolumn} 
\usepackage{color}
\usepackage[colorlinks=true,urlcolor={blue},citecolor={blue}]{hyperref}
\usepackage{ulem}

\usepackage[utf8]{inputenc}

\usepackage{amsmath}
\usepackage{amsfonts}
\usepackage{amssymb}

\usepackage{graphicx}

\usepackage{hyperref}
\hypersetup{
    colorlinks=true,       
    linkcolor=blue,          
    citecolor=blue,        
    filecolor=blue,      
    urlcolor=blue           
}

\begin{document}
\title{ Strong anharmonicity induces quantum melting of charge density wave in 2H-NbSe$_2$ under pressure}

\author{Maxime Leroux}
\email{mleroux@anl.gov}
\altaffiliation{Present address : Materials Science Division, Argonne National Laboratory, 9700 S. Cass Avenue, Argonne, Illinois 60439,
USA}
\affiliation{Universit\'e Grenoble Alpes, CNRS, I. Néel, F-38000 Grenoble, France}
\author{Ion Errea}
\affiliation{Donostia International Physics Center (DIPC), Manuel de Lardizabal Pasealekua 4, 20018 Donostia-San Sebasti\'an, Basque Country, Spain}
\affiliation{IKERBASQUE, Basque Foundation for Science, 48011, Bilbao, Spain}
\affiliation{IMPMC, UMR CNRS 7590, Sorbonne Universit\'es - UPMC Univ. Paris 06,  MNHN, IRD, 4 Place Jussieu, F-75005 Paris, France}
\author{Mathieu Le~Tacon}
\affiliation{Max-Planck-Institut~f\"{u}r~Festk\"{o}rperforschung, Heisenbergstr.~1, D-70569 Stuttgart, Germany}
\author{Sofia-Michaela Souliou}
\affiliation{Max-Planck-Institut~f\"{u}r~Festk\"{o}rperforschung, Heisenbergstr.~1, D-70569 Stuttgart, Germany}
\author{Gaston Garbarino}
\affiliation{European Synchrotron Radiation Facility - Grenoble, France}
\author{Laurent Cario}
\affiliation{Institut des Mat\'eriaux Jean Rouxel (IMN), Universit\'e de Nantes, CNRS, 2 rue de la Houssini\`ere, BP3229, 44322 Nantes, France}
\author{Alexey Bosak}
\affiliation{European Synchrotron Radiation Facility - Grenoble, France}
\author{Francesco Mauri}
\affiliation{IMPMC, UMR CNRS 7590, Sorbonne Universit\'es - UPMC Univ. Paris 06,  MNHN, IRD, 4 Place Jussieu, F-75005 Paris, France}
\author{Matteo Calandra}
\email[]{matteo.calandra@upmc.fr}
\affiliation{IMPMC, UMR CNRS 7590, Sorbonne Universit\'es - UPMC Univ. Paris 06,  MNHN, IRD, 4 Place Jussieu, F-75005 Paris, France}
\author{Pierre Rodi\`ere}
\email{pierre.rodiere@neel.cnrs.fr}
\affiliation{Universit\'e Grenoble Alpes, CNRS, I. Néel, F-38000 Grenoble, France}

\date{\today}

\begin{abstract}

The pressure and temperature dependence of the phonon dispersion of 2H-NbSe$_2$ is measured by inelastic X-ray scattering. A strong temperature dependent soft phonon mode, reminiscent of the Charge Density Wave (CDW), is found to persist up to a pressure as high as 16~GPa, far above the critical pressure at which the CDW disappears at 0~K. By using ab initio calculations beyond the harmonic approximation, we obtain an accurate, quantitative, description of the (P,T) dependence of the phonon spectrum. Our results show that the rapid destruction of the CDW under pressure is related to the zero mode vibrations - or quantum fluctuations - of
the lattice renormalized by the anharmonic part of the lattice potential.  
The calculations also show that the low-energy longitudinal acoustic mode that drives the CDW transition barely contributes to superconductivity, explaining the insensitivity of the superconducting critical temperature to the CDW transition.

\end{abstract}
\maketitle

The interplay between charge density wave order, i.e. a static modulation of the electronic density close to the Fermi level, and superconductivity has attracted much attention \cite{Ghiringhelli_Science2012, LeTacon2013}. This is the central issue of a long standing debate in simple transition metal dichalcogenides without strong electronic correlations, such as 2H-NbSe$_2$ \cite{TMD_INS_Moncton, Wilson2001}. At $T_{CDW}$=33.5 K, 2H-NbSe$_2$ undergoes a second order phase transition towards an incommensurate CDW phase which coexists with superconductivity below $T_c$ = 7.2 K.
The anisotropy and temperature dependence of the electronic band structure has been widely studied by ARPES and STM measurements \cite{Yokoya2001, Valla2004,Kiss2007,Guillamon08,Borissenko2009, Rahn2012,Arguello2015}. The CDW is coupled to a periodic lattice distortion through a strong electron-phonon coupling. The transition is associated with a softening of a longitudinal acoustic phonon mode as temperature is lowered to $T_{CDW}$~\cite{Weber_NbSe2}.  
It was already noticed that the high order phonon fluctuations and strong electron-phonon interactions explain some of the key features of the formation of the CDW in this system \cite{Inglesfield80,Varma_PRL1983}.  In the specific case of NbSe$_2$ at ambient pressure, the reduction of the phonon lifetime through the electron-phonon coupling is a key ingredient on the temperature dependence of the phonon dispersion \cite{Flicker2015}.

\begin{figure}[h]
	\centering
	\includegraphics[width=0.9\columnwidth]{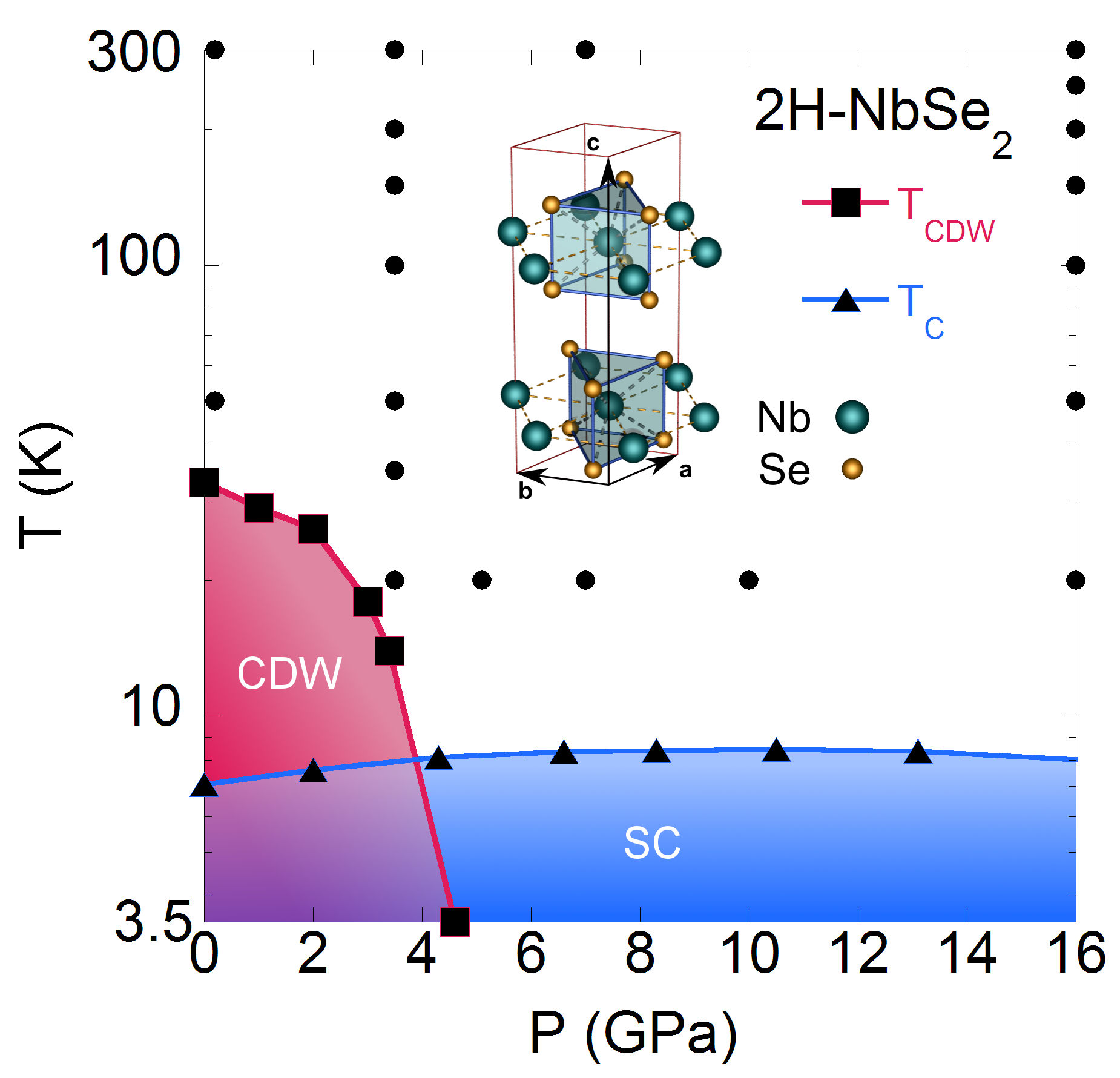}
	\caption{{\bf Phase diagram of NbSe$_2$. Squares indicate the limit of the CDW phase measured 
         in Ref.~\cite{Berthier1976_SSC} with the addition of the point at 4.6\,GPa 
 from Ref.~\cite{Feng08052012}, 
         triangles indicate the superconducting $T_c$ measured 
         in \cite{Suderow_nbse2_pression}, and circles indicate the 
         pressure and temperature of our IXS measurements. Inset : crystallographic structure of 2H-NbSe$_2$ \cite{TMD_INS_Moncton}. }}
	\label{fig:0}
\end{figure}

At ambient pressure, published ab initio calculations reproduce successfully the lattice instability on a wide part of the Brillouin zone around the experimentally observed
${\bf q}_{\rm CDW}$ ~\cite{PhysRevB.80.241108, Weber_NbSe2,Weber2013}. However, these calculations are carried out at the harmonic level, and therefore at T=0K. The temperature dependence can be qualitatively reproduced using a Gaussian smearing of the Fermi surface which reduces the contribution of the electron-phonon interaction. Quantitatively however, the temperatures that would effectively yield this smearing are unphysical, as they exceed the experimentally observed one by several orders of magnitude.

The quantitative failure of the harmonic ab-initio calculations is even worse under pressure. Experimentally, by applying an hydrostatic pressure, $T_{CDW}$ is driven to 0~K and the CDW instability disappears above the critical pressure $P_{CDW}$(T = 3.5 K)=4.6~GPa~\cite{Feng08052012}. Around this quantum phase transition superconductivity remains essentially unaffected, as shown on the (T,P) phase diagram in Fig. 1~\cite{Berthier1976_SSC,Suderow_nbse2_pression}. However, harmonic ab initio calculations fail to describe the ground state of NbSe$_2$ as a function of pressure (see SI and ~\cite{0953-8984-21-39-395502}). Indeed, they  predict (see Fig.~\ref{fig:3} and SI) a CDW instability up to $P_C=14\,$GPa, a pressure that largely exceeds the experimentally observed one. This  persistence of the CDW instability in the harmonic calculations is reminiscent of the effect of iso-valent and iso-electronic substitution of Se with lighter S. The harmonic calculation also predicts a CDW instability for NbS$_2$ that is not observed experimentally \cite{NbS2_IXS_Leroux}.

Taken together, these facts indicate that for an accurate description of the ground state (even at T=0K) of these systems, calculations beyond the standard harmonic approximation must be carried out. The impact of anharmonicity has recently been highlighted in different systems. To mention only a few, the thermodynamical properties of the metal-insulator transition in strongly correlated VO$_2$~\cite{Budai14}, the transport properties of thermoelectric PbTe~\cite{Delaire11} or the electron-phonon interaction above the Verwey transition in magnetite Fe$_3$O$_4$~\cite{Hoesch_PRL2013} are all strongly influenced by anharmonicity. Even at low temperatures, where its effects are expected to be weaker, it must be taken into account to explain the pressure dependence of the crystal structure of Calcium~\cite{Errea_PRL2011}, the inverse isotope effect in superconducting PdH~\cite{Errea_PRL2013}, or the phonon spectra at the proximity of the liquid-solid transition of Helium at 0K~\cite{Glyde}. Finally, anharmonicity is crucial for understanding the ground state of a system when a quantum electronic phase transition coupled to the lattice occurs, as in e.g. ferroelectrics~\cite{Rowley14} or CDW compounds~\cite{Gruner_DW,Varma_PRL1983, NbS2_IXS_Leroux, Weber_NbSe2}. In most cases however, these effects are computationally too expensive to be evaluated in standard first-principle calculations, and therefore often disregarded.

Here, we report the investigation of the temperature and pressure dependence of the phonon spectra of NbSe$_2$ by using high resolution inelastic X-ray scattering (IXS) from a crystal in a diamond anvil cell, up to pressures as large as 16~GPa at T = 20~K. We observe that the strong temperature dependence of the soft phonon mode is reduced by pressure, but still present even at 16~GPa. 

We show that the temperature and pressure dependence of the phonon spectra can be accurately accounted for when the effects of anharmonicity are explicitly included in the calculation.
Anharmonic effects were calculated within the newly developed stochastic self-consistent harmonic approximation (SSCHA), yielding an unprecedented agreement with the experimentally determined temperature dependence of the phonon spectra at various pressures.
We show that at low temperatures, for pressures between $P_{CDW}$ and $P_C$, the CDW is destroyed by quantum fluctuations, which easily overcome the double-well lattice potential and, thus, are strongly affected by the anharmonic part of the potential. Surprisingly, we observe that anharmonic effects dominate the low-energy phonon spectrum of 2H-NbSe$_2$ in a large pressure range, extending at least up to 16 GPa $>> P_{CDW}$ and down to the lowest temperatures.
Finally, we demonstrate that a large electron-phonon interaction, mostly due to optical modes rather than to the soft longitudinal acoustic mode which drives the CDW instability, contributes substantially to superconductivity in 2H-NbSe$_2$, and naturally explains its insensitivity to the occurrence of a CDW.

Single crystals of 2H-NbSe$_2$ were grown using the vapor growth technique in sealed quartz tubes with iodine as a transport agent~\cite{TMD_INS_Moncton}. Typical dimensions of the crystals are $100\times 100 \times 50\,\mathrm{\mu m^3}$ ($\vec{a}\times\vec{b}\times\vec{c}$).
Pressure was generated using a diamond anvil cell, loaded with helium as a pressure transmitting medium to ensure highly hydrostatic conditions. Temperature was lowered using a custom-designed $^4$He cryostat, and pressure was varied \textit{in-situ} at low temperature using a helium-pressurized membrane. Each cell contained two rubies to monitor the pressure \textit{in-situ} using the fluorescence technique. Measurements of the phonon dispersion of 2H-NbSe$_2$ were taken between room temperature and 20\,K, and for pressures ranging from 0.2 to 16\,GPa.

IXS measurements were carried out on beamline ID28 at the ESRF. The x-ray beam was aligned along the c-axis of the crystal. We used the (9,9,9) reflection on the backscattering monochromator with an incident energy of 17.794\,keV, and a corresponding energy resolution of 1.3\,meV HWHM (lorentzian fit of the elastic peak). The incident beam was focused by a multilayer mirror into a spot of $100 \times 60 \: \mathrm{\mu m}$ (width $\times$ height). We used $20 \times 60 \: \mathrm{mm}$ (width $\times$ height) slits at a 7\,m distance giving a momentum resolution of $0.014\,$\AA$^{-1}$ ($a^* \times c^* \approx 2.1 \times 0.50\,$\AA$^{-1}$) in the (H,0,L) plane, and $0.042\,$\AA$^{-1}$ ($b^*\times \sin (60\!^{\circ})\approx 1.8 \,$\AA$^{-1}$) in the perpendicular direction.
We measured the phonon dispersion along the $\mathrm{\Gamma M}$ direction close to $\Gamma_{200}$ (\textit{i. e.} (2-h, 0, 0)), where both longitudinal optical and acoustic like soft phonon branches can be observed. According to the balance of the dynamical structure factor of the 2 phonon modes and so of the inelastic scattering function S($\bf q$,$\omega$), the observation of one of the two branches can be selected. For each transfer wave vector investigated, an energy scan up to 15meV was performed. 

All the spectra were fitted using the standard Levenberg-Marquadt algorithm, with the following free parameters : position and amplitude of the elastic peak; position, linewidth and amplitude of the phonon; and a constant background. We make the standard assumptions~\cite{Fok,Burkel_IXS} that the elastic peak is Lorentzian, and that the phonon lineshape can be modeled by a damped harmonic oscillator convoluted by a normalized Lorentzian with a width corresponding to the resolution of the detector.


\textit{Ab initio} calculations were performed within density functional perturbation theory and the generalized gradient approximation~\cite{PhysRevLett.77.3865} making use of the {\sc  Quantum ESPRESSO}~\cite {0953-8984-21-39-395502} code. An ultrasoft pseudopotential (norm-conserving) for Nb (Se), a 35\,Ry energy cutoff, and a 24$\times$24$\times$8 mesh for the electronic integrations were used. An Hermitian-Gaussian smearing of 0.01 Ry was used. Harmonic dynamical matrices were calculated within linear response in
a grid of
6$\times$6$\times$4 $\mathbf{q}$ points.  The small smearing parameter and the dense electronic mesh used are needed to converge the phonon frequency of the longitudinal acoustic mode at the CDW wave vector (cf. SI). The stochastic self-consistent harmonic approximation (SSCHA)~\cite{PhysRevB.89.064302,Errea_PRL2013}
calculations were performed using a 3$\times$3$\times$1 supercell,
yielding
anharmonic dynamical matrices in the commensurate 3$\times$3$\times$1
grid.
The difference between the SSCHA dynamical matrices and the harmonic
dynamical matrices in the 3$\times$3$\times$1 grid was Fourier
interpolated to the points of the finer 6$\times$6$\times$4 grid.
Adding the harmonic dynamical matrices in the 6$\times$6$\times$4 grid to
the result of the interpolation, the SSCHA dynamical matrices
in the finer 6$\times$6$\times$4 mesh were obtained.
The experimental lattice parameters\cite{Feng08052012} with relaxed internal coordinates were used because in these conditions the results weakly depend on the exchange-correlation potential chosen.

The electron-phonon calculations were performed by using maximally
localized Wannier functions~\cite{PhysRevB.56.12847,PhysRevB.65.035109}
and interpolation
of the electron-phonon matrix elements as in Ref.~\cite{PhysRevB.82.165111}.
We used 14 Wannier functions (d-states of Nb and Se p$_z$ states) and
a $6\times6\times4$ electron-momentum grid. We used the Fourier interpolated
SSCHA dynamical matrices in the electron-phonon coupling calculations.
The average electron-phonon coupling was interpolated using electron and phonon momentum
grids of $24\times24\times 8$ randomly displaced from the origin.

\begin{figure}
\includegraphics[width=1.0\columnwidth]{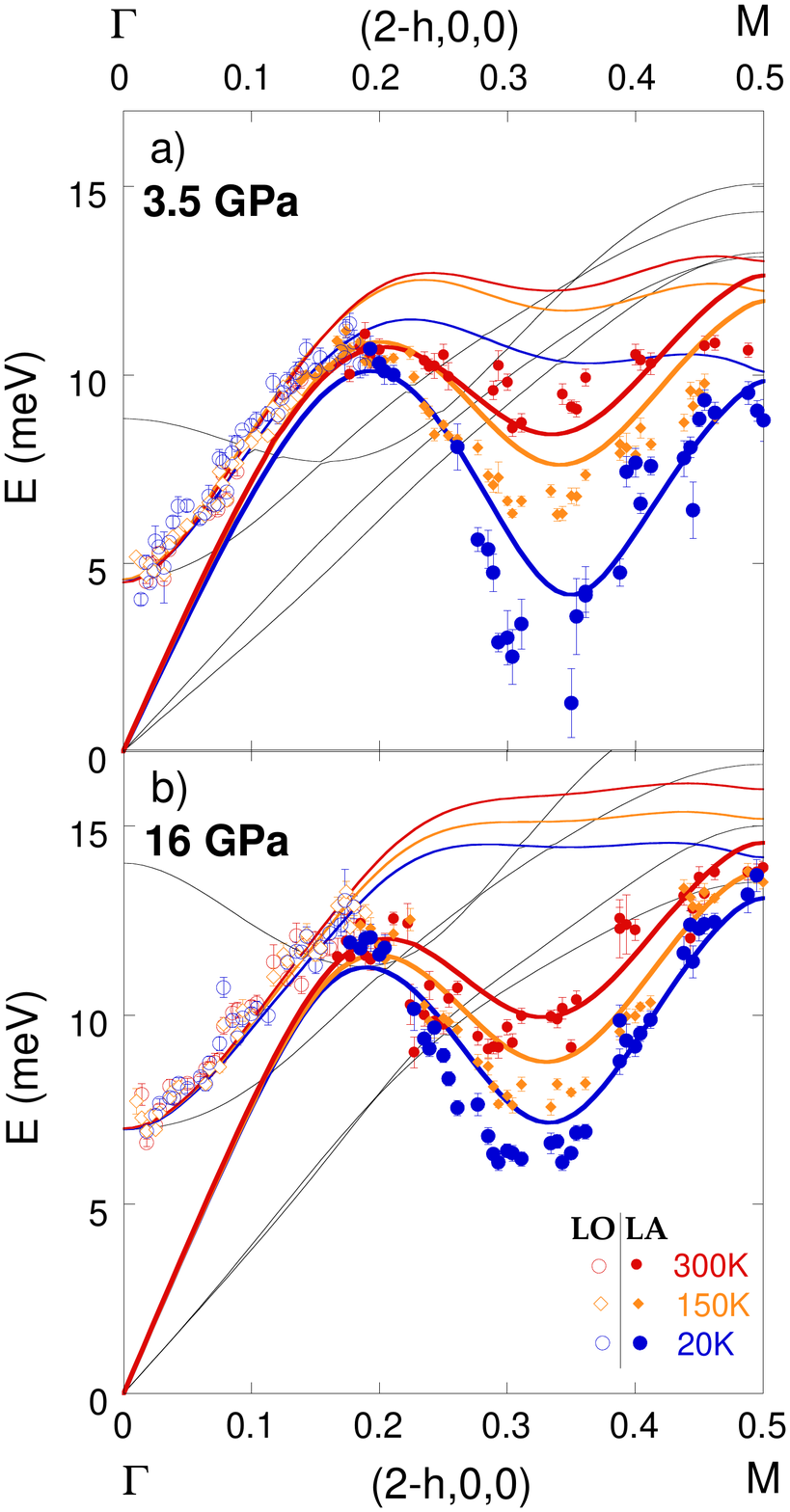}
	\caption{\bf{    Phonon dispersion
                      at 3.5\,GPa (a) and 16\,GPa (b).
                        Closed (resp. open) dots represent the experimentally determined longitudinal acoustic (longitudinal optic) soft phonon dispersions, and lines represent the results of phonon calculations.  Close to the $\Gamma$ point, the phonon dispersion shows clearly an optical branch. Closer to the M point, the energy of the phonon observed correspond to the acoustic like branch calculated.
                         Colored thick and thin lines represent the SSCHA results
                         for the longitudinal acoustic and optical anharmonic phonons, respectively,
                         at several temperatures.
                         The thin black lines represent
                         other phonon modes present in this energy
                         range calculated within the SSCHA at 20 K, which are
                         very harmonic and barely depend on temperature (see SI).}
}
	\label{fig:2}
\end{figure}


\begin{figure}
    \includegraphics[width=1.\columnwidth , bb = 50 550 400 842] {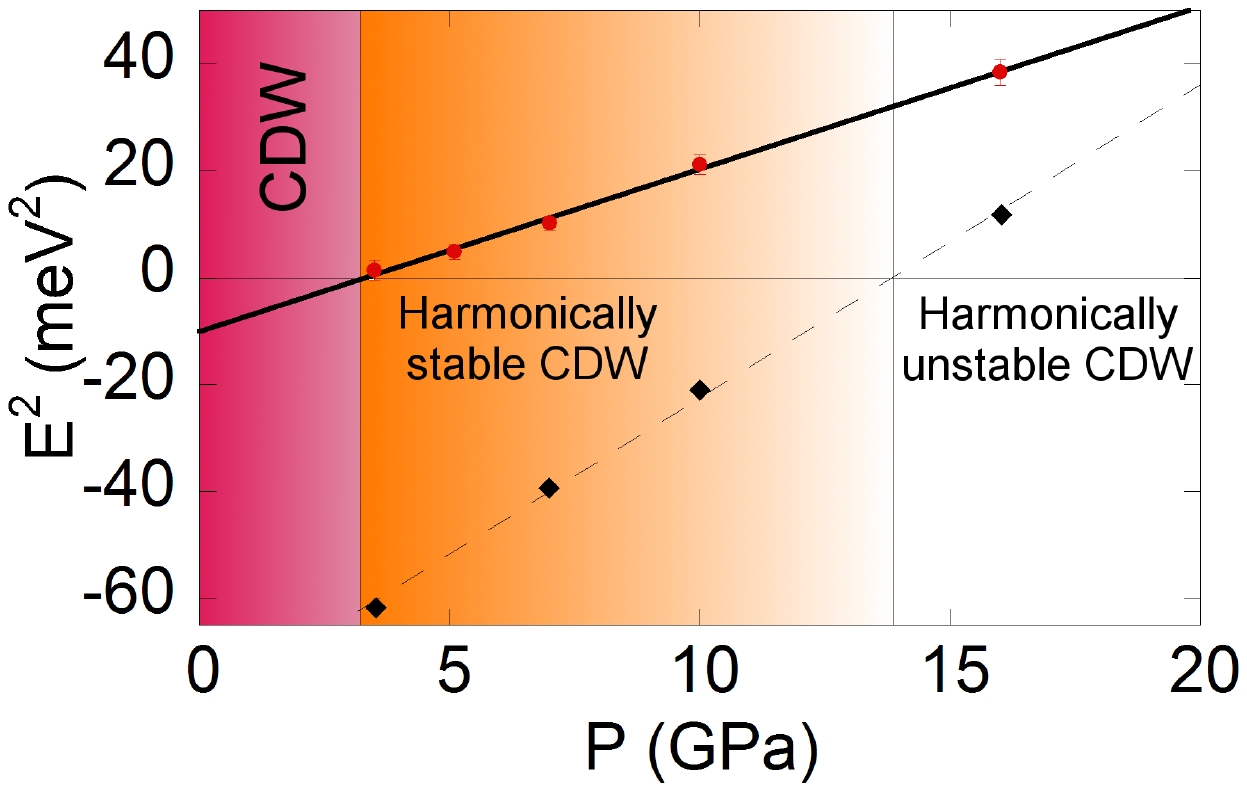}
    \caption{{\bf Pressure dependence of the square of the measured energy of the longitudinal acoustic soft phonon
at the CDW ordering wavevector and at 20\,K (red dots) compared to the square of the energy calculated in the harmonic approximation (black diamonds). Lines are linear fits. The red region indicates the range of pressure where a CDW is observed experimentally at 20\,K.
The orange region indicates the range of pressure where harmonic calculations predict a CDW ground state.}}
    \label{fig:3}
\end{figure}

The temperature dependence of the low energy phonon dispersions recorded at pressures of 3.5 and 16 GPa are shown in Fig. \ref{fig:2} (some of the corresponding raw data is shown in the supplementary materials).
At $T~=20$ K and $P=3.5$ GPa, the system is very close to the quantum critical regime reported in ref.~\onlinecite{Feng08052012}. No particular enhancement of the elastic line intensity close to the CDW wavevector ${\bf q}_{\rm CDW}\approx (1.67,0,0)$ is observed. On the other hand, a colossal temperature dependent softening of the longitudinal acoustic phonon mode is seen : from $\sim$ 9 to $\sim$ 1.5 meV between room temperature and 20 K. While this mode is expected to condense around $\sim$ 12 K~\cite{Feng08052012}, this temperature could not be reached at this pressure with the current experimental set-up. As we increase pressure to 16 GPa, the amplitude of the phonon softening is significantly reduced, but remains sizeable as the mode's energy almost decreases by a factor of 2 from 300 to $20$ K.  So far, most of the theoretical and experimental work on NbSe$_2$ has focused on the interplay between the electronic susceptibility and the electron-phonon interaction as possible mechanisms for the CDW formation \cite{PhysRevB.73.205102, Zhu2015}. The large temperature dependence indicates that strong anharmonic effects survive over a pressure range that is way larger than the experimentally reported CDW stability range ($0 < P < P_{CDW}$) and might therefore play an important role that was up to now largely disregarded.

In Fig. \ref{fig:3}, we plot the square of the soft phonon frequency measured at T = 20 K at ${\bf q}_{\rm CDW}$ as a function of pressure. The linear dependence indicates that the mode hardens as $\sqrt{P-P_{CDW}}$. We also plot the square of the phonon frequency calculated within the harmonic approximation, which is negative - i.e. indicating that the system is unstable against CDW formation - up to 14 GPa $>> P_{CDW}$. This is reminiscent of the situation encountered at ambient pressure in NbS$_2$, where we have observed a large phonon softening upon cooling, insufficient however to induce the CDW formation despite the predictions of the harmonic ab-initio calculation~\cite{NbS2_IXS_Leroux}.

To quantify the effect of anharmonicity and check whether it is responsible for the destabilization of the CDW between $P_{CDW}$ and 14 GPa, we have carried out a series of calculations for the pressures and temperatures experimentally investigated, within the SSCHA approach~\cite{Errea_PRL2013, PhysRevB.89.064302} (see SI) which allows us to access
directly the anharmonic free energy of the system, with full inclusion of the anharmonic potential terms.

The results are superimposed to our experimental dispersions in Fig.\ref{fig:2}. Not only is the instability suppressed as experimentally observed, but the agreement between the measured and calculated dispersions over the entire pressure range is remarkable compared to harmonic ab-initio calculation. A small overestimation of phonon calculation is nevertheless observed in the range around ${\bf q}_{\rm CDW}$. 

A key result here is that, even at low temperature, anharmonic terms play a key role in the destruction of the harmonically stable long-range CDW order. This is surprising at first glance since anharmonic effects usually manifest themselves at finite temperature, through the effect of many-body interactions between thermally excited phonons. These account for the lattice thermal expansion and for the decay of the phonons lifetime at high temperatures, but are usually strongly suppressed at low temperatures.
On the other hand, given that atoms vibrate even at zero temperature due to the zero-point fluctuations, their motion can in principle be affected 
by the anharmonic potential. This appears to be the case in this system where, below 14 GPa, the harmonic part of the potential that drives the lattice instability is overwhelmed by the anharmonic part. The balance between these two parts of the potential tells whether there is a lattice instability (driven by the CDW) or not. This bears stark analogy with the destruction of long range magnetic order from spin-spin quantum fluctuations in low dimensional quantum magnets~\cite{Balents}.

Finally we note that these many-body effects are relevant not only at ${\bf q}={\bf q}_{\rm CDW}$, but extend up to the M point, the boundary of the Brillouin zone at ${\bf q}={\bf a^*}/2$,  where two low-energy modes, the longitudinal acoustic mode that drives the CDW transition and an optical mode, are substantially hardened with respect to the result of the harmonic calculation.

\begin{figure}
\includegraphics[width=0.9\columnwidth]{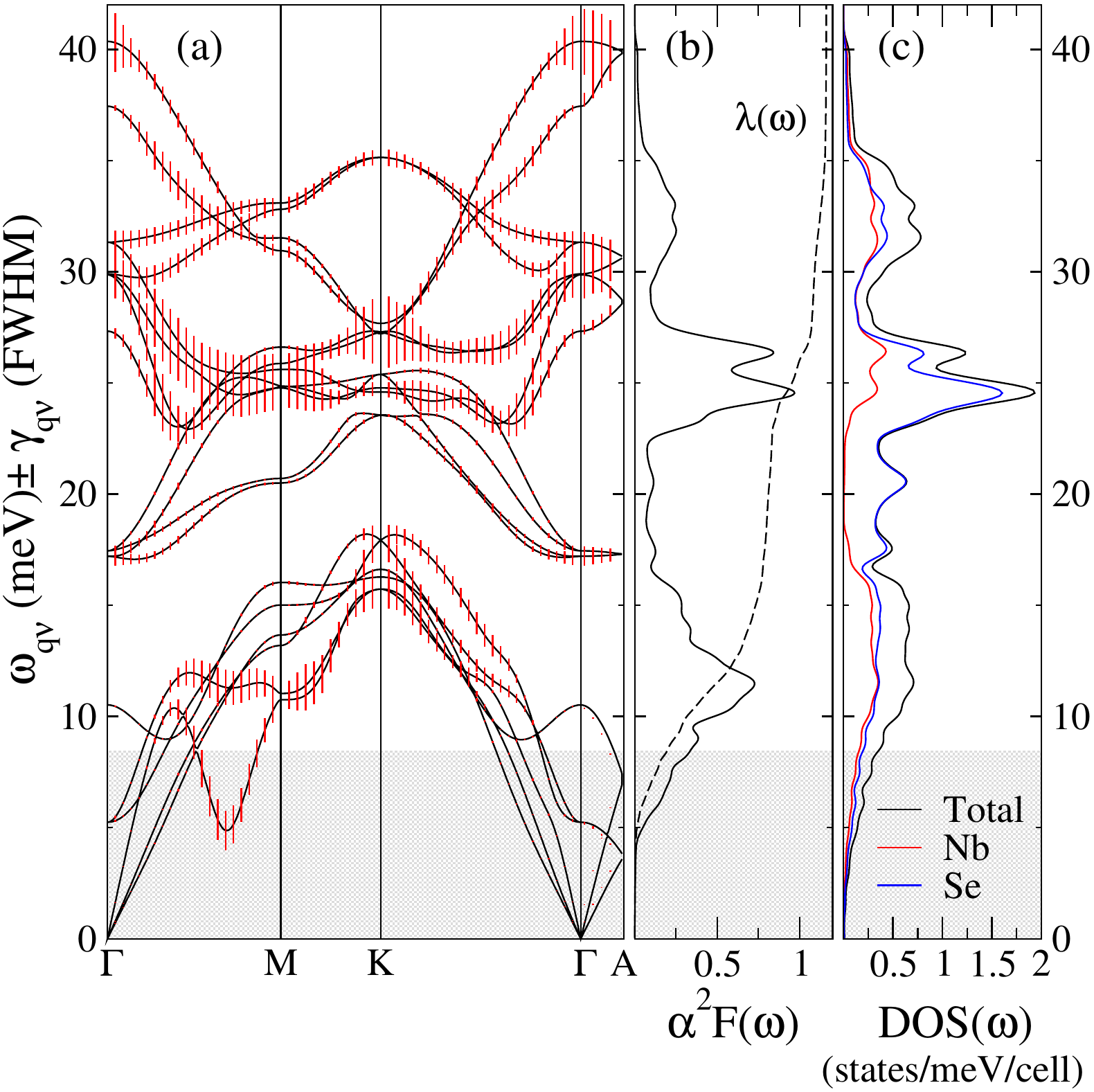}
    \caption{{\bf (a) Calculated phonon dispersion at 7 GPa and 0 K with the SSCHA.
The size of the bars is proportional to the electron-phonon
contribution to the phonon linewidth (the bar is twice the FWHM).
(b) The Eliashberg function $\alpha^2F(\omega)$ and the
integrated electron-phonon coupling $\lambda(\omega)$.(c) The phonon density of states decomposed  into the different atoms. In all figures the grey background denotes the integration area used to infer that the soft acoustic longitudinal mode contributes 17\% of the total electron-phonon coupling.}}
    \label{fig:4}
\end{figure}


The importance of the electron-phonon interaction in the CDW formation has been emphasized~\cite{Weber_NbSe2,PhysRevB.73.205102}, but its role in superconductivity remains debated, as is, more generally, the interplay between these two orders. Indeed, the absence of reliable ab initio phonon calculations in the high symmetry phase, has prevented a calculation of the superconducting critical temperature.
Having achieved a detailed understanding of the low energy part of the experimental phonon spectra and having shown how anharmonicity can suppress the long range ordered CDW phase, we can now address the superconducting properties of NbSe$_2$ by calculating the Eliashberg function and the integrated electron-phonon coupling (see SI) as a function of pressure.
As of now, this calculation can only be done in the high pressure phase, where the number of atoms per unit cell is low enough. We show in Fig. \ref{fig:4}
the results of the calculation at $7$ GPa (data for other pressures are shown in the SI).
We find that the average electron-phonon coupling is as large as
$\lambda=1.16$ (see Tab. \ref{tab:Tc} for other pressures),
demonstrating that NbSe$_2$ is a strong coupling superconductor.

\begin{table}[t]
\begin{tabular}{c | c  c  c  c }
\hline \hline
P (GPa) & $\lambda$ & $\omega_{log}$ (meV) & $T_c$ calc. (K) & $T_c$ exp. (K)  \\
\hline
3.5           & 1.28 & 12.6 & 11.3 & 7.8 \\
7             & 1.16 & 13.7 & 10.5 & 8.2 \\
16            & 0.91 & 16.4 & 7.8  & 7.8 \\
\hline
\hline
\end{tabular}
\caption{Calculated $\lambda$, logarithmic frequency average $\omega_{log}$, and  $T_c$ values for 2H-NbSe$_2$ at 3.5, 7 and 16 GPa.
The McMillan equation with $\mu^{*}=0.16$ is used to calculate T$_c$.
Experimental data are taken from Ref.~\cite{Suderow_nbse2_pression}.}
\label{tab:Tc}
\end{table}

Previous works~\cite{PhysRevB.80.241108,Gorkov_2012,Weber_NbSe2,Weber2013,Arguello2015}
identified the large electron-phonon coupling of the soft acoustic mode
at ${\bf q}_{{\rm CDW}}$ as the mechanism responsible for the CDW formation.
For this reason, it is natural to expect that this soft mode
also plays a key role in superconductivity. However, its contribution to the average electron-phonon coupling is
less than $17\%$ (as it can be inferred
from the integrated $\lambda(\omega)$ in Fig. \ref{fig:4}),
meaning that it has only a marginal role in superconductivity. Indeed,
the strong coupling
of this mode is very localized around ${\bf q}_{{\rm CDW}}$, and
its contribution to $\lambda$ averages out when integrating over the Brillouin zone.
The modes sustaining superconductivity are given by
the two main contributions to the Eliashberg function: a broad
peak centred around $\approx 11.5$ meV and a second one in the
$22-26$ meV region. These two
features contribute to $\approx 70\%$ of the total electron-phonon
coupling. Further insights are obtained by decomposing the phonon density of states (DOS)
in atomic vibrations along different directions. As the DOS and
$\alpha^2F(\omega)$ are very similar, the decomposition will also
apply to the Eliashberg function.
The low energy feature at $11.5$ meV is mainly due to Nb and Se vibrations
parallel to the Nb plane, corresponding to the flat phonon band along
$\Gamma$M suffering from a large anharmonic correction,
while the features in the $22-26$ meV range are mostly attributed to
out-of-plane and in-plane Se displacements with a non negligible Nb
component. In both modes sustaining superconductivity
there is a substantial Se component, upturning the conventional wisdom
that in superconducting dichalcogenides the chalcogene is accessory
while the transition metal plays a key role \cite{Wilson2001}. Our work demonstrates that
the superconducting properties strongly depend on both the transition
metal and the chalcogene, offering a natural explanation for the different relation between CDW and
superconductivity encountered in different dichalcogenides \cite{Tissen, Sipos, Kusmartseva}.
Moreover, as optical phonon modes contribute strongly to superconductivity in 2H-NbSe$_2$, while the CDW is determined by the softening of the longitudinal acoustic mode,
the superconducting $T_c$ is insensitive to the presence of the CDW,
explaining the phase diagram in Fig. \ref{fig:0}.

Our work points toward the key role of anharmonicity in destroying the CDW order not only through the usual thermal phonon-fluctuations, but also through quantum phonon-fluctuations, and on a very wide pressure range above the quantum phase transition. Our results, in this respect, are fundamental and relevant for a very large set of materials exhibiting a CDW instability - or more generally a second order electronic quantum phase transition coupled to the lattice - such as transition metal dichalcogenides, cuprates, Bechgaard salts and transition metal bronzes.

\section*{Acknowledgements}
Sabrina Salmon-Bourmand is thanked for her help in the synthesis of 2H-NbSe$_2$ single crystals. P.R. and M.L. acknowledge financial support from  the French National Research Agency through Grant No. ANR-12-JS04-0003-01 SUBRISSYME.
M.C, F.M. and I.E. acknowledge support from the Basque
Government (Grant No. BFI-2011-65), Spanish Ministry of Economy and Competitiveness (FIS2013-48286-C2-2-P,) Graphene Flagship and
ANR (contract ANR-13-IS10-0003-01) and Prace (2014102310).
Calculations were performed at IDRIS, CINES,
DIPC and at CEA TGCC.


{}





\end{document}